\documentclass[a4paper,11pt]{article}

\usepackage{jcappub}
\usepackage{amssymb}
\usepackage{graphicx}
\usepackage{amsmath}
\usepackage{mathrsfs}
\usepackage{hyperref}
\usepackage{subfigure}

\title{\boldmath Scalar-induced gravitational waves including isocurvature perturbations with lattice simulations}

\author[a,b]{Xiang-Xi Zeng}
\emailAdd{zengxiangxi@itp.ac.cn}

\affiliation[a]{Institute of Theoretical Physics, Chinese Academy of Sciences (CAS), Beijing 100190, China}
\affiliation[b]{School of Physical Sciences, University of Chinese Academy of Sciences (UCAS), Beijing 100049, China}

\abstract{Scalar-induced gravitational waves (SIGWs) open a unique window into early-universe physics. While their generation from adiabatic perturbations has been extensively studied, the contribution from isocurvature perturbations remains largely unexplored. In this work, we develop a lattice simulation framework to compute the stochastic gravitational wave background from both pure isocurvature and mixed initial conditions. Our numerical results show excellent agreement with semi-analytical predictions in the pure isocurvature case. We further analyze multi-peak structures under general initial conditions and find that they closely match those produced in purely adiabatic scenarios. Additionally, we examine SIGWs in early matter-dominated eras, revealing that the peak amplitude and spectral slope are sensitive to the microphysical properties of the dominant field, such as the primordial black hole mass, abundance, or soliton decay rate. This study establishes lattice simulations as a robust tool for predicting SIGW spectra from complex primordial perturbations, with important implications for interpreting current and future gravitational wave observations.}

\begin{document}
\maketitle
\flushbottom

\section{Introduction}
\label{sec:intro}
A series of recent results from Pulsar Timing Array (PTA) collaborations (NANOGrav~\cite{NANOGrav:2023gor}, EPTA~\cite{EPTA:2023fyk}, PPTA~\cite{Reardon:2023gzh}, and CPTA~\cite{Xu:2023wog}) have brought us to the era of detecting a stochastic gravitational wave background (SGWB). The upcoming of advanced gravitational wave detectors, from space-based missions like LISA~\cite{LISA:2017pwj}, Taiji~\cite{Hu:2017mde, Ruan:2018tsw}, TianQin~\cite{TianQin:2015yph} as well as the more advanced DECIGO~\cite{Kawamura:2006up} and BBO~\cite{Corbin:2005ny} to future ground-based facilities such as Einstein Telescope (ET)~\cite{ET:2019dnz} and Cosmic Explorer (CE)~\cite{CE:2019iox}), will open a window onto a landscape of SGWBs encoding a wealth of new physics (see Refs.~\cite{Caprini:2018mtu, Cai:2017cbj, Bian:2021ini} for reviews). Beyond astrophysical sources, the SGWB offers a unique probe of the early universe, potentially encoding information from epochs far beyond the reach of traditional cosmic microwave background (CMB) observations. Among the most promising mechanisms for generating such a background are scalar-induced gravitational waves (SIGWs)(see Ref.~\cite{Domenech:2021ztg} for a review), which are produced when primordial scalar perturbations re-enter the horizon and source tensor fluctuations through second-order gravitational interactions~\cite{Ananda:2006af, Baumann:2007zm, Kohri:2018awv, Espinosa:2018eve, Terada:2025cto}.

The properties of SIGWs are intimately tied to the nature of the primordial perturbations. While the majority of studies have focused on adiabatic perturbations~\cite{Cai:2018dig, Unal:2018yaa, Adshead:2021hnm, Inomata:2018epa, Cai:2024dya}, consistent with the models of single field inflation and strongly constrained by CMB observations~\cite{Planck:2018jri, WMAP:2003ivt, Planck:2018vyg}, another fundamental class, isocurvature perturbations~\cite{Kodama:1984ziu, Kodama:1986fg, Kodama:1986ud, Langlois:2003fq, Bucher:1999re}, remains a compelling and less explored possibility (see Ref.~\cite{Domenech:2023jve} for a review, and see Refs.~\cite{Yu:2023jrs, Luo:2025lgr, Yu:2025jgx} for some recent works).  Isocurvature modes, characterizing relative density contrasts between different cosmological components (e.g., radiation and dark matter), are a generic prediction of many multi-field inflation models~\cite{Chung:2017uzc, Chung:2021lfg, Lyth:2001nq, Polarski:1994rz, Linde:1985yf}, phase transition~\cite{Buckley:2024nen} and could be associated with the formation of primordial black holes (PBHs)~\cite{Inman:2019wvr, Papanikolaou:2020qtd, Domenech:2020ssp, Domenech:2021wkk, Domenech:2024wao, Balaji:2024hpu, Papanikolaou:2022chm, Sasaki:2025zao} or other exotic non-topological relics like Q-balls~\cite{Lozanov:2023aez, Lozanov:2023knf}. Crucially, isocurvature perturbations, particularly those of Poissonian origin such as from PBHs, are inherently large, reaching amplitudes of $O(1)$~\cite{ Papanikolaou:2020qtd, Domenech:2020ssp}. Consequently, they can generate a substantial SIGW signal and must be carefully accounted for.

The exact semi-analytical formalism for isocurvature-induced SIGWs was first given in Ref.~\cite{Domenech:2021and}, which is strictly valid only under the condition that the relevant perturbation modes enter the horizon long before matter-radiation equality, while the complexity escalates significantly for mixed initial conditions (with both adiabatic and isocurvature perturbations). In contrast, lattice simulations offer a robust alternative, with an intrinsic advantage in accurately capturing effects from non-Gaussianity. While this method has previously been applied exclusively to adiabatic fluctuations~\cite{Zeng:2025cer}, we systematically extend it in this work to incorporate isocurvature perturbations and mixed initial conditions. We note in Ref.~\cite{Fernandez:2023ddy}, the authors have applied a hybrid N-body and lattice simulation to study the GWs from structure formation in the early matter-dominated era, while our method has treated the matter as a fluid field, since the effects of structure formation are not our focus.

An early matter-dominated (eMD) epoch represents another compelling scenario~\cite{Assadullahi:2009nf}. During eMD, the gravitational potential is stabilized, leading to a significant enhancement of SIGWs~\cite{Inomata:2019ivs, Inomata:2020lmk}. A critical factor is the transition timescale out of eMD: a fast transition triggers a resonant amplification (poltergeist mechanism)~\cite{Inomata:2020lmk}, while a slow transition suppresses the signal~\cite{Inomata:2019zqy, Pearce:2023kxp, Zeng:2025ecx}. Despite existing numerical work on adiabatic initial conditions~\cite{Inomata:2019zqy, Pearce:2023kxp, Zeng:2025ecx}, little attention has been paid to the numerical investigation of isocurvature perturbations. For this purpose, we adopt the lattice simulation method, which is well-suited to this problem.

The remainder of this paper is structured as follows. Section 2 is devoted to the theoretical formalism for cosmological perturbations incorporating isocurvature modes, along with a description of the numerical methodology. A comprehensive analysis comparing lattice simulations with semi-analytical results and the multi-peak phenomena under general initial conditions is presented in Section 3. Section 4 investigates the implications of this method within various eMD era scenarios. The paper concludes in Section 5 with a summary of our results and a discussion of prospective future work. We work in reduced Planck units where $c=\hbar=1$ and $M_{\mathrm{pl}} = (8\pi G)^{-1/2}=1$. All our simulations employ a lattice size of $N=256$, with the initial time set to $\tau_i = 4/N$.

\section{Evolution of curvature and isocurvature perturbations}\label{sec:SIGW}

Isocurvature perturbations are generated by entropy perturbations, which necessitate the presence of multiple constituents in the universe~\cite{Kodama:1984ziu, Kodama:1986fg, Kodama:1986ud}. Accordingly, our analysis requires at least two distinct fluids. Here, as a typical example, we assume the universe is composed of radiation and non-relativistic matter, and generalize to other kinds of fluids similar to Ref.~\cite{Domenech:2025ffb} or the cases with more than two components should be straightforward. The energy-momentum tensors of the radiation and the matter are, respectively, given by
\begin{align}
    T_{(r)\mu\nu} &= (\rho_{(r)} + p_{(r)})u_{(r)\mu}u_{(r)\nu} + p_{(r)} g_{\mu\nu},\\
    T_{(m)\mu\nu} &= \rho_{(m)} u_{(m)\mu}u_{(m)\nu},
\end{align}
where $g_{\mu\nu}$ is the metric, $\rho$ denotes the energy density, $p$ is the pressure, and $u_{\mu}$ is the four-velocity of fluid. Hereafter, we mainly follow the conventions established in Ref.~\cite{Domenech:2023jve}, and the metric in Newtonian gauge is expressed as
\begin{align}
    \mathrm{d}s^2 = a^2(\tau) \left[ -(1+2\Psi)\mathrm{d}\tau^2 + \left((1+2\Phi)\delta_{ij} + h_{ij} \right)\mathrm{d}x^i\mathrm{d}x^j   \right],
\end{align}
where $a$ is the scale factor, $\tau$ is the conformal time, $\Psi$ and $\Phi$ denotes the scalar potential, and $h_{ij}$ are the tensor perturbations. Given the definition of $h_{ij}$, the energy spectrum of GWs is expressed as~\cite{Maggiore:1999vm}
\begin{align}
    \Omega_{\mathrm{GW}}(\tau) = \frac{\rho_{\mathrm{GW}}(\tau)}{\rho_c(\tau)}=\frac{1}{12\mathcal{H}^2}\langle h_{ij}^{\prime}h_{ij}^{\prime} \rangle,
\end{align}
with $\rho_c(\tau)$ denoting the critical energy density at time $\tau$ and angle brackets describing a volume average.

\subsection{Equations of motion}\label{subsec:eom}
At the level of background, one has the Friedmann equations
\begin{align}
    3\mathcal{H}^2 &= a^2(\rho_{(m)} + \rho_{(r)})\label{eq:a},\\
    3(\mathcal{H}^2 + 2\mathcal{H}^{\prime}) &= -a^2\rho_{(r)}\label{eq:dH},
\end{align}
where $\mathcal{H}$ is the conformal Hubble parameter, defined as $a^{\prime}/a$, with the prime denoting differentiation with respect to conformal time $\tau$. For the energy-momentum conservation equations, an energy transfer $Q$ between components is typically accounted for
\begin{align}
    \nabla_{\mu} T^{(m)\mu\nu}=& -Q^{\nu},\\
    \nabla_{\mu} T^{(r)\mu\nu}=& Q^{\nu}.
\end{align}
Given our focus on the early matter-dominated era (eMD), we accordingly assume that $Q^{\mu}$ originates entirely from the matter sector, then, $Q^{\mu}$ can only be proportional to the four-velocity of the matter, thus, we can define it as
\begin{align}
    Q^{\mu} \equiv Qu_{(m)}^{\mu}.
\end{align}
For the PBH-dominated case, $Q$ should be proportional to the number denisty of PBHs $n_{\mathrm{PBH}} = \rho_{\mathrm{PBH}}/M_{\mathrm{PBH}}$ and their mass decay rate, which leads to
\begin{align}
    Q = - n_{\mathrm{PBH}}u_{(m)}^{\mu} \partial_{\mu}M_{\mathrm{PBH}}.
\end{align}
Hence, the above equations at the zeroth order arrive at
\begin{align}
    \rho_{(m)}^{\prime} + 3\mathcal{H}\rho_{(m)} &= -aQ,\label{eq:rhom}\\
    \rho_{(r)}^{\prime} + 4\mathcal{H}\rho_{(r)} &= aQ.\label{eq:rhor}
\end{align}
At first order in perturbation theory, we neglect the transverse modes for $u_{(m)},u_{(r)}$ and define $u_{(m), i}\equiv a\partial_i v_{(m)}$ and $u_{(r), i}\equiv a\partial_i v_{(r)}$. Assuming the absence of anisotropic stress, $\Phi=-\Psi$ follows. To derive the perturbative equations, we have used the \textbf{xPand}~\cite{Pitrou:2013hga} to perform the calculation. The perturbed Einstein equations are given by
\begin{align}
    &6\mathcal{H}\Phi^{\prime} + 6\mathcal{H}^2\Phi - 2\Delta\Phi = a^2 (\delta\rho_{(m)} + \delta\rho_{(r)}) \equiv a^2\delta \rho,\label{eq:drho} \\
    &\Phi^{\prime} + \mathcal{H}\Phi = \frac{1}{2}a^2\left(\rho_{(m)} v_{(m)} + \frac{4}{3}\rho_{(r)} v_{(r)} \right) \equiv \frac{1}{2}a^2 \rho V, \label{eq:V}\\
    &\Phi^{\prime\prime} + 3\mathcal{H}\Phi^{\prime} + (\mathcal{H}^2 + 2\mathcal{H}^{\prime})\Phi = -\frac{1}{6}a^2\delta\rho_{(r)},\label{eq:phi} 
\end{align}
while the energy-momentum conservation yields
\begin{align}
    &\delta\rho_{(m)}^{\prime} + 3\mathcal{H}\delta\rho_{(m)} + \rho_{(m)}(3\Phi^{\prime} + \Delta v_{(m)}) = -a\delta Q + a\Phi Q,\label{eq:drhom} \\
    &\delta\rho_{(r)}^{\prime} + 4\mathcal{H}\delta\rho_{(r)} + \frac{4}{3}\rho_{(r)}(3\Phi^{\prime} + \Delta v_{(r)}) = a\delta Q -a\Phi Q,\label{eq:drhor} \\
    &v_{(m)}^{\prime} + \mathcal{H}v_{(m)} - \Phi = 0, \label{eq:vm}\\
    &v_{(r)}^{\prime} + \frac{1}{4}\frac{\delta\rho_{(r)}}{\rho_{(r)}} - \Phi = \frac{aQ}{\rho_{(r)}} \left( \frac{3}{4}v_{(m)} - v_{(r)} \right). \label{eq:vr}
\end{align}
At second order in perturbation theory, one can find the equations of tensor modes
\begin{align}
    h_{ij}^{\prime\prime} + 2\mathcal{H}h_{ij}^{\prime} - \Delta h_{ij} = T_{ij}^{~~lm}S_{lm}, \label{eq:hij}
\end{align}
with the source term
\begin{align}
    S_{ij} =& 4\partial_i \Phi \partial_j \Phi + 2a^2 \left( \rho_{(m)}\partial_i v_{(m)} \partial_j v_{(m)} + \frac{4}{3}\rho_{(r)} \partial_i v_{(r)} \partial_j v_{(r)} \right)\nonumber\\
    =& 4\partial_i \Phi \partial_j \Phi + 6c_s^2\frac{\rho}{\rho_r}\partial_i\left(\frac{\Phi^\prime}{\mathcal{H}}+\Phi\right)\partial_j\left(\frac{\Phi^\prime}{\mathcal{H}}+\Phi\right) + 
    6a^2c_s^2\rho_m\partial_i V_{\mathrm{rel}}\partial_j V_{\mathrm{rel}},
\end{align}
where $V_{\mathrm{rel}} \equiv v_{(m)}-v_{(r)}$ and $T_{ij}^{~~lm}$ is the transverse-traceless projection operator defined by
\begin{align}
    T_{ijlm} = P_{il}P_{jm} - \frac{1}{2}P_{ij}P_{lm},~~~~P_{ij} = \delta_{ij} - \frac{\partial_i\partial_j}{\partial^2}.
\end{align}
It is important to note that the term involving $V_{\mathrm{rel}}$ has been neglected in many of the previous literature, even though it can have a non-negligible effect in certain scenarios, as highlighted in Ref.~\cite{Kumar:2024hsi}. In this work, we have explicitly tested its impact in Appendix~\ref{app:vrel} and find that it remains negligible in the PBH-dominated scenario discussed in Section~\ref{sec:pbh}. This result is consistent with the analysis presented in Ref.~\cite{Domenech:2020ssp}. In our simulation,  the scale factor $a(\tau)$ and conformal Hubble parameter $\mathcal{H}$ are evolved using Eq.~\eqref{eq:a}, while Eq.~\eqref{eq:dH} determines $\mathcal{H}^{\prime}$. The energy densities $\rho_{(m)}$ and $\rho_{(r)}$ are advanced via Eqs.~\eqref{eq:rhom} and \eqref{eq:rhor}, respectively. For the perturbative sector, Eqs.~\eqref{eq:phi}, \eqref{eq:drhom}, \eqref{eq:drhor}, \eqref{eq:vm}, \eqref{eq:vr}, and \eqref{eq:hij} govern the evolution of $\Phi,\delta\rho_{(m)},\delta\rho_{(r)},v_{(m)},v_{(r)}$ and $h_{ij}$, respectively. Simulations are performed in a comoving cubic box with periodic boundary conditions, following Ref.~\cite{Zeng:2025cer}. Spatial derivatives are computed using a Fourier pseudospectral method, and time integration is handled by a fourth-order Runge-Kutta scheme.

\subsection{The initial conditions}

It's conventional to define the isocurvature perturbation $S$ to describe the behavior of relative density fluctuations between different components
\begin{align}
    S \equiv \frac{\delta\rho_{(m)}}{\rho_{(m)}} - \frac{\delta\rho_{(r)}}{\rho_{(r)} + p_{(r)}} = \frac{\delta\rho_{(m)}}{\rho_{(m)}} - \frac34\frac{\delta\rho_{(r)}}{\rho_{(r)}},
\end{align}
where we used $p_{(r)} = 1/3\rho_{(r)}$. From the equations of motion in Subsection~\ref{subsec:eom}, the evolution equation for $S$ can be derived by
\begin{align}
    S^{\prime} = -\Delta V_{\mathrm{rel}} - a(\delta Q - Q\Phi)\frac{3}{4}\frac{\rho_{(m)} + 4\rho_{(r)}/3}{\rho_{(m)}\rho_{(r)}} + aQ\left(\frac{\delta\rho_{(m)}}{\rho_{(m)}^2} + \frac{3}{4}\frac{\delta\rho_{(r)}}{\rho_{(r)}^2} \right)
\end{align}
with $V_{\mathrm{rel}}\equiv v_{(m)}-v_{(r)}$ denoting the relative velocity, and the equations of $V_{\mathrm{rel}}$ and $\Phi$ can also be obtained and rewritten as
\begin{align}
    &V_{\mathrm{rel}}^{\prime} + 3c_s^2\mathcal{H} V_{\mathrm{rel}} + \frac{3}{2a^2\rho_{(r)}}c_s^2\Delta\Phi + \frac{3\rho_{(m)}}{4\rho_{(r)}}c_s^2S -
    \frac{aQ}{4\rho_{(r)}}\frac{\rho V - 4(\rho_{(m)}+\rho_{(r)})V_{\mathrm{rel}}}{\rho_{(m)} + 4\rho_{(r)}/3} = 0,\label{eq:vrel} \\
    &\Phi^{\prime\prime} + 3\mathcal{H}(1+c_s^2)\Phi^{\prime} + \left(\mathcal{H}^2(1+3c_s^2) + 2\mathcal{H}^{\prime}\right)\Phi - 
    c_s^2\Delta\Phi = \frac{a^2}{2}\rho_{(m)} c_s^2S,\label{eq:phis}
\end{align}
where the sound speed is defined as
\begin{align}
    c_s^2 = \frac{1}{3}\left(1 + \frac{3\rho_{(m)}}{4\rho_{(r)}} \right)^{-1} = \frac{4}{9}\frac{\rho_{(r)}}{\rho_{(m)} + 4\rho_{(r)}/3}.
\end{align}
When there is no energy transfer, namely, $Q=\delta Q=0$, the scale factor has an analytical form
\begin{align}\label{eq:ana_a}
    \frac{a(\tau)}{a_{\mathrm{eq}}} = 2\left(\frac{\tau}{\tau_*}\right) + \left(\frac{\tau}{\tau_*}\right)^2,
\end{align}
where $a_{\mathrm{eq}}$ denotes $a(\tau_{\mathrm{eq}})$ at the time of matter-radiation equality, and $\tau_{\mathrm{eq}}=(\sqrt{2}-1)\tau_{*}$. Using this solution, one can derive the initial conditions for the energy densities $\rho_{(m)}$ and $\rho_{(r)}$. An exact solution of $\Phi$ and $S$ can also be derived on superhorizon scale~\cite{Kodama:1986ud, Domenech:2023jve} without energy transfer
\begin{align}
    S(\xi, \vec{x}) &= S_i(\vec{x}),\label{eq:Sols} \\
    \Phi(\xi, \vec{x}) &= \Phi_{i,\mathrm{ad}}(\vec{x})\left(\frac{8}{5\xi^3}(\sqrt{1+\xi} -1) -\frac{4}{5\xi^2} + \frac{1}{5\xi} + \frac{9}{10} \right)\nonumber \\
    & + S_i(\vec{x})\left(\frac{16}{5\xi^3}(1-\sqrt{1+\xi}) +\frac{8}{5\xi^2} - \frac{2}{5\xi} + \frac{1}{5} \right),\label{eq:solphi}
\end{align}
where $\xi\equiv a/a_{\mathrm{eq}}$, and the solution of their derivative can be directly calculated
\begin{align}
    S^{\prime}(\xi,\vec{x}) &= 0, \\
    \Phi^{\prime}(\xi, \vec{x}) &= f(\tau)\Phi_{i,\mathrm{ad}}(\vec{x})\left(\frac{8}{5\xi^3} -\frac{1}{5\xi^2} + \frac{4}{5\xi^3\sqrt{1+\xi}} - \frac{24(\sqrt{1+\xi}-1)}{5\xi^4} \right)\nonumber \\
    & + f(\tau)S_i(\vec{x})\left(-\frac{16}{5\xi^3} +\frac{2}{5\xi^2} - \frac{8}{5\xi^3\sqrt{1+\xi}} - \frac{48(1-\sqrt{1+\xi})}{5\xi^4} \right), \\
    f(\tau) &= 2\left(\frac{1}{\tau_*} + \frac{\tau}{\tau_*^2}\right). \nonumber
\end{align}
The cosmological Gaussian random field $f_g$ is usually defined in momentum space~\cite{Mukhanov:2005sc}. Adopting the Fourier transfer convention
\begin{align}
    f_g(\vec{x}) = \int \frac{\mathrm{d}^3 x}{(2\pi)^3} f_g(\vec{k}) e^{i \vec{k} \cdot \vec{x}},
\end{align}
and assuming the Fourier modes are complex Gaussian variables with a probability density function
\begin{align}
    \mathscr{P}[f_g(\vec{k})] = \frac{1}{\pi\sigma_k^2} \exp \left\{ -\frac{|f_g(\vec{k})|^2}{\sigma_k^2}  \right\},
\end{align}
leads to 
\begin{align}
    \langle f_g(\vec{k}) f_g(\vec{q}) \rangle = \sigma_k^2 \delta^3(\vec{k} + \vec{q}), 
\end{align}
thus, we have $\sigma_k^2 = (2\pi)^3\times(k^3/(2\pi^2))\mathcal{P}_{f_g}(k)$, where $\mathcal{P}_{f_g}(k)$ is the dimensionless power spectrum of $f_g(\vec{k})$. Therefore, after defining the dimensionless power spectrum of Gaussian random fields $\Phi_g$ and $S_g$\footnote{Generalizing to non-Gaussian fields is straightforward~\cite{Zeng:2025cer}.}
\begin{align}
    \langle \Phi_g(\vec{k})\Phi_g(\vec{q}) \rangle &= (2\pi)^3\times\frac{2\pi^2}{k^3}\mathcal{P}_{\Phi_g}(k)\delta^3(\vec{k}+\vec{q})\nonumber \\
    &= 
    (2\pi)^3\times\frac{2\pi^2}{k^3} \left(\frac{5+3\omega}{3+3\omega}\right)^2 \mathcal{P}_{\zeta_g}(k)\delta^3(\vec{k}+\vec{q}), \\
    \langle S_g(\vec{k})S_g(\vec{q}) \rangle &= (2\pi)^3\times\frac{2\pi^2}{k^3}\mathcal{P}_{S_g}(k)\delta^3(\vec{k}+\vec{q}),
\end{align}
one can obtain the initial value of $\Phi_{i,\mathrm{ad}}$ (``ad'' means adiabatic) and $S_i$~\cite{Zeng:2025cer}, where the equation of state parameter $w$ is defined as $\omega=p_{(r)}/(\rho_{(r)}+\rho_{(m)})$ and $\zeta_g$ is the gauge invariant curvature perturbation. For the initial relative velocity $V_{\mathrm{rel}}$, combining Eqs.~\eqref{eq:vrel}, \eqref{eq:phis}, and the exact solutions~\eqref{eq:Sols} and~\eqref{eq:solphi} shows that during radiation-domination or matter-domination,
\begin{align}
    V_{\mathrm{rel}}^{\prime} + 3c_s^2\mathcal{H}V_{\mathrm{rel}}\simeq 0,
\end{align}
which leads to the regular solution $V_{\mathrm{rel}}=0$. Then, Eq.~\eqref{eq:V} can be applied to obtain the initial value of $v_{(m)}$ and $v_{(r)}$, while Eq.~\eqref{eq:drho} and $S_i$ can be used to derive the initial conditions for $\delta\rho_{(m)}$ and $\delta\rho_{(r)}$. The initial conditions for $h_{ij}$ are set as usual $(h_{ij})_i = 0$.

\section{Simulation results without energy transfer}
As a first demonstration, we initially compare our simulation results with semi-analytical predictions in the pure isocurvature case. We subsequently generalize the initial conditions to include both adiabatic and isocurvature perturbations, and present illustrative plots highlighting their characteristic peak structures.

\subsection{Comparing with the semi-analytical predictions}
The exact semi-analytical formulas in the isocurvature case were first derived in Ref.~\cite{Domenech:2021and}. Provided the fluctuations enter the horizon significantly prior to matter-radiation equality, the energy spectrum of induced GWs at time $\tau_c$ can be expressed as
\begin{align}
    \Omega_{\mathrm{GW}}(k,\tau_c) = \frac{2}{3}\int_{0}^{\infty} \mathrm{d}v \int_{|1-v|}^{1+v}\mathrm{d}u \left(\frac{4v^2 -(1-u^2+v^2)^2}{4uv}\right)^2\overline{I^2(k, \tau_c, u, v)}\mathcal{P}_S(ku)\mathcal{P}_S(kv),
\end{align}
where $\overline{I^2(k, \tau_c, u, v)}$ is the kernel function, which can be found in Appendix~\ref{app:iso}. For concreteness, we consider a Gaussian bump power spectrum of both curvature perturbation $\zeta$ and isocurvature perturbation $S$
\begin{align}
    &\mathcal{P}_\zeta (k) = A\frac{(k/k_{*1})^3}{\sqrt{2\pi}\Delta_{\mathrm{ad}}}\exp \left[-\frac{(k/k_{*1}-1)^2}{2\Delta_{\mathrm{ad}}^2} \right],\label{eq:psad} \\
    &\mathcal{P}_S (k) = B\frac{(k/k_{*2})^3}{\sqrt{2\pi}\Delta_{\mathrm{iso}}}\exp \left[-\frac{(k/k_{*2}-1)^2}{2\Delta_{\mathrm{iso}}^2} \right],\label{eq:psiso}
\end{align}

\begin{figure}
    \centering
    \includegraphics[width=0.49\linewidth]{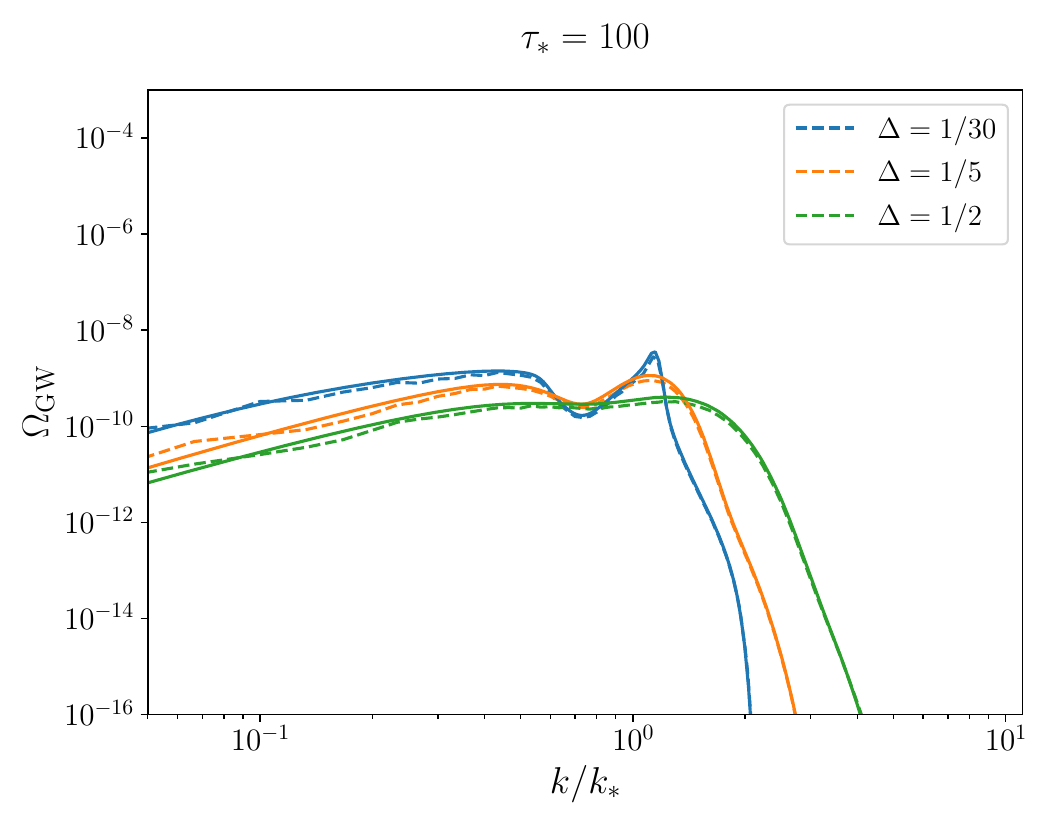}
    \includegraphics[width=0.49\linewidth]{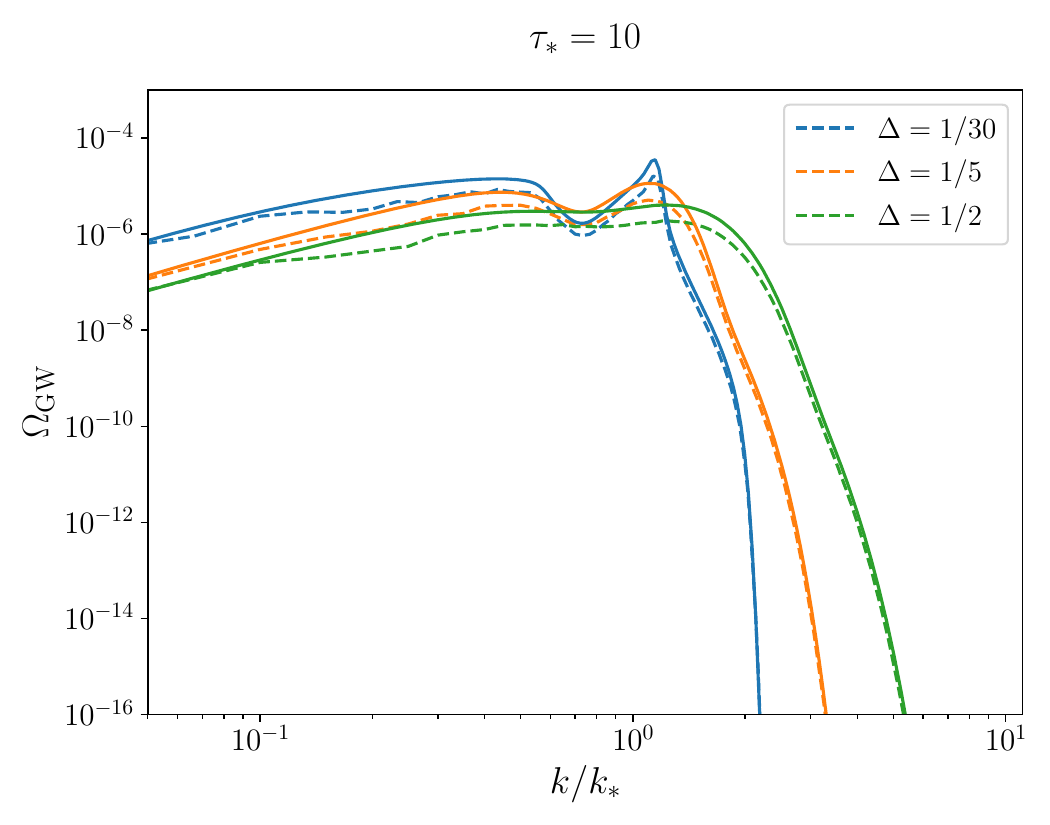}
    \caption{This figure compares the energy spectra of gravitational waves obtained via lattice simulation and semi-analytical methods. Solid lines correspond to semi-analytical results, and dashed curves represent lattice simulation outcomes. The initial amplitude of the isocurvature perturbation is $B=100$, while the peak scale is chosen as $k_*=30$. On the left panel, the matter-radiation equality parameter from Eq.~\eqref{eq:ana_a} is set to $\tau_* = 100$; on the right panel, $\tau_* = 10$.}
    \label{fig:anyiso}
\end{figure}

Fig.~\ref{fig:anyiso} compares the energy spectra of gravitational waves from lattice simulations and semi-analytical methods, showing good agreement between each other. As noted in Ref.~\cite{Domenech:2021and}, the semi-analytical formulas remain accurate only when the characteristic modes enter the horizon well before matter-radiation equality; otherwise, modifications are required. The right panel of Fig.~\ref{fig:anyiso} demonstrates a suppression in the amplitude of the GW energy spectra when the peak modes enter the horizon near matter-radiation equality.

\subsection{The peak structures of general initial conditions}\label{sec:peak}
The multi-peak structure in the GW energy spectrum, originating from multi-peaked curvature perturbations, was initially investigated in Ref.~\cite{Cai:2019amo} and later extended to scenarios with non-Gaussianity in Ref.~\cite{Zeng:2024ovg}. However, these studies focus exclusively on adiabatic perturbations. A similar multi-peak structure is also expected for isocurvature perturbations and mixed initial conditions. Considering a multi-peak Dirac delta power spectrum
\begin{align}\label{eq:psdelta}
    \mathcal{P}_{S}(k) = \sum_{i=1}^{n}B_i \delta(\ln(k/k_{i})). 
\end{align}
Here we fix $0<k_1<k_2<...<k_n$ without loss of generality and define $\tilde{k}_i\equiv k/k_i$ for later convenience. One can easily obtain the multi-peak structure for the initial isocurvature perturbation
\begin{align}\label{eq:multi}
    \Omega_{\mathrm{GW,iso}} = \frac{2}{3}\sum_{i,j}^{n}B_iB_j\tilde{k}_i^{-1}\tilde{k}_j^{-1}&\left(\frac{4\tilde{k}_i^{-2} - (1-\tilde{k}_j^{-2} + \tilde{k}_i^{-2})^2}{4\tilde{k}_i^{-1}\tilde{k}_j^{-1}}\right)^2\overline{I^2(k,\tau_c\rightarrow\infty,\tilde{k}_i^{-1},\tilde{k}_j^{-1})}\nonumber\\
    &\Theta(k_i+k_j-k)\Theta(k - |k_i-k_j|),
\end{align}
which indicates that isocurvature perturbations will exhibit a peak structure similar to that of adiabatic perturbations. Consequently, mixed initial conditions are also expected to yield an identical peak profile. The peak arises from the logarithmic pole in the kernel function $\overline{I^2}$ and hence emerges precisely when 
\begin{align}\label{eq:peak}
    \left(\frac{k_i}{k}+\frac{k_j}{k} \right) = \sqrt{3}.
\end{align}
The number of such peaks is governed by momentum conservation, as represented by the step function in Eq.~\eqref{eq:multi}, which requires
\begin{align}
    |k_i - k_j| \leq k \leq k_i + k_j.
\end{align}
Assuming $k_j=nk_{i}$ and $n>1$, the appearance of an additional peak requires $(1+n)/\sqrt{3}k_i \geq (n-1)k_i$, which yields the constraint $n\leq 2+\sqrt{3}$. However, in the isocurvature case, a suppression factor of the form $(k_{\mathrm{eq}}/k_i)^4$ or $(k_{\mathrm{eq}}/k_i)^2(k_{\mathrm{eq}}/k_j)^2$ arises. This implies that certain peaks in the mixed initial condition exhibit an explicit dependence on $k_{\mathrm{eq}}$.

\begin{figure}
    \centering
    \includegraphics[width=0.49\linewidth]{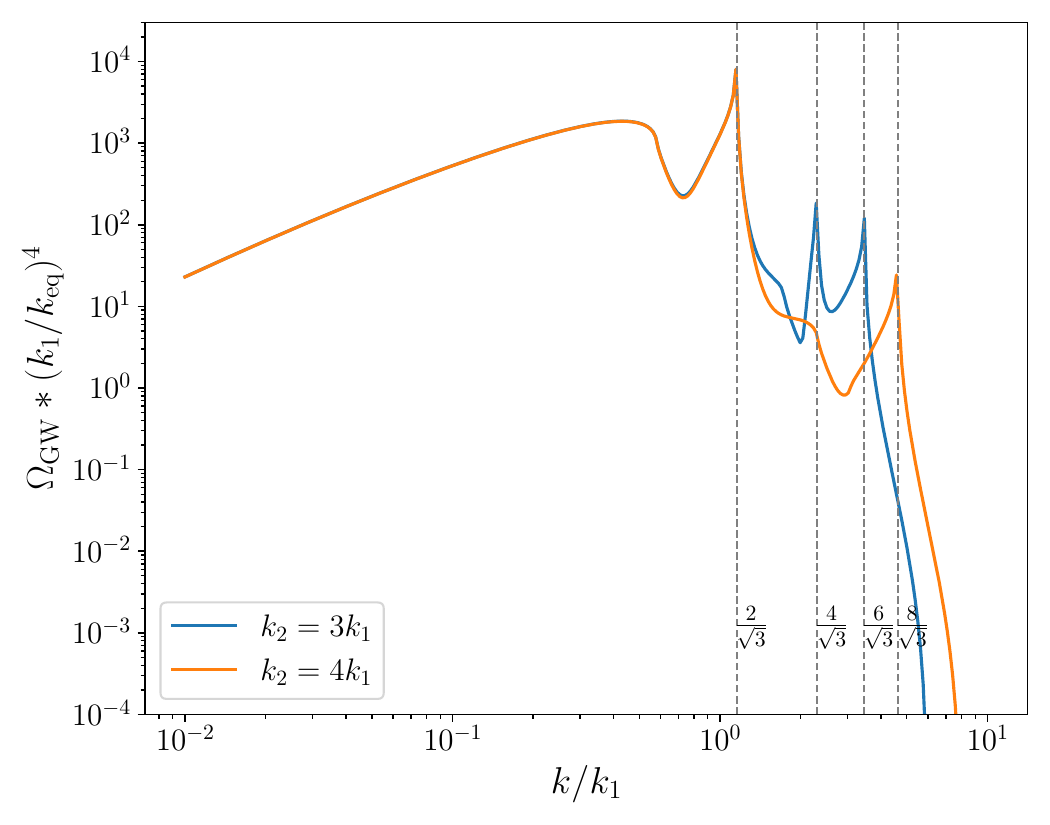}
    \includegraphics[width=0.49\linewidth]{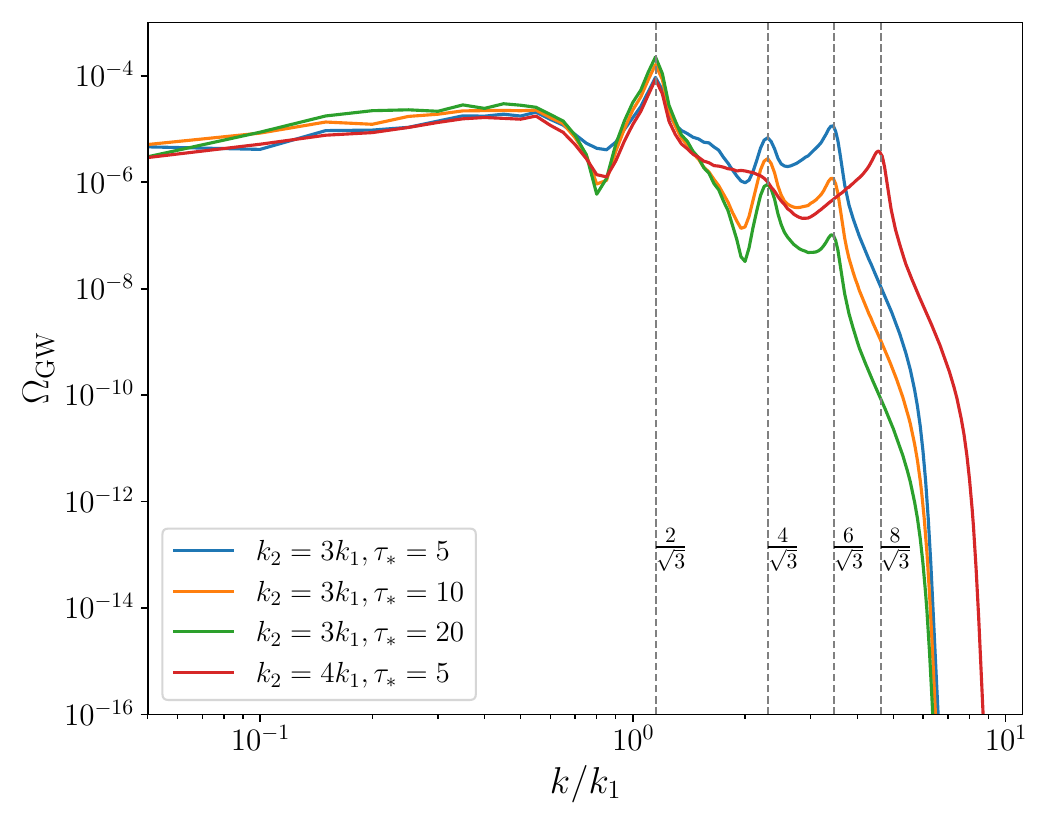}
    \caption{Peak structures for the isocurvature case (\textit{left}) and the mixed initial condition case(\textit{right}). \textit{Left}: Semi-analytical results for the peak structure in the pure isocurvature scenario. Two representative cases are presented: $k_2 = 3k_1$ (three distinct peaks) and $k_2 = 4k_1$ (two peaks). \textit{Right}:  Lattice simulation results for the mixed initial condition, shown for different values of $\tau_*$ and $k_2/k_1$. Here, $k_1=20$ denotes the peak scale of the adiabatic perturbation, and $k_2$ the peak scale of the isocurvature perturbation. The width of power spectrum has been set as $\Delta_{\mathrm{ad}} = \Delta_{\mathrm{iso}}=1/30$.}
    \label{fig:multi}
\end{figure}
\begin{figure}
    \centering
    \includegraphics[width=0.49\linewidth]{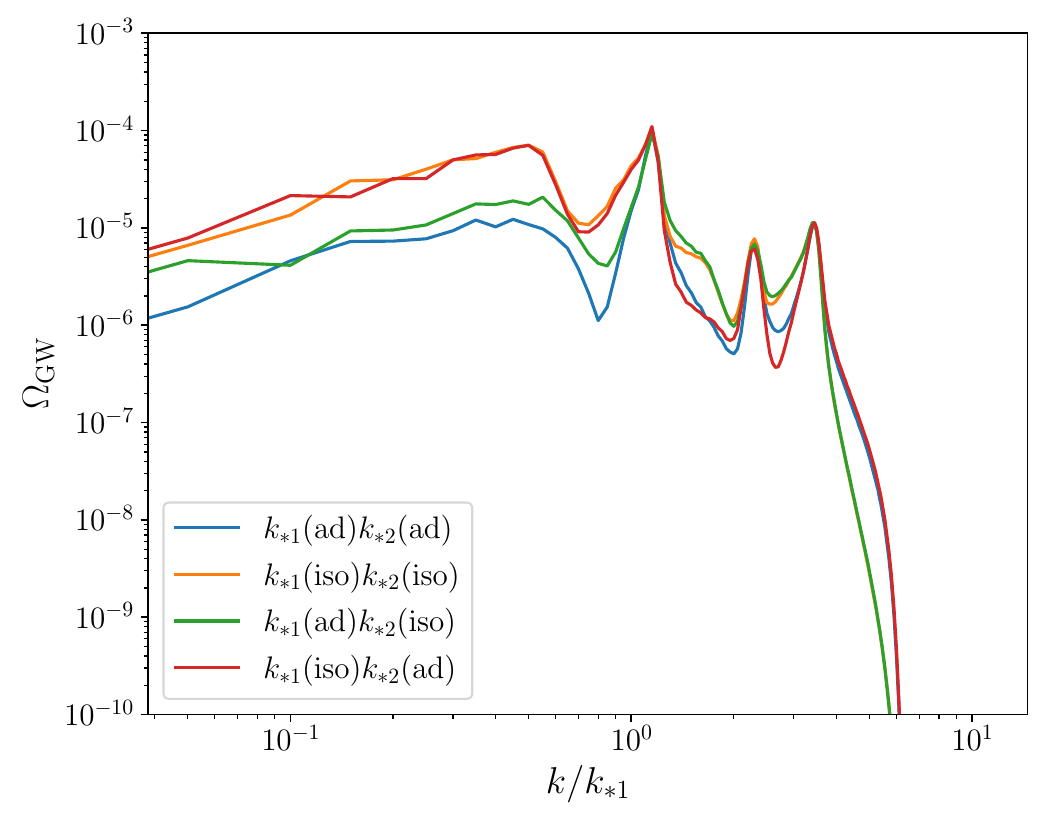}
    \includegraphics[width=0.49\linewidth]{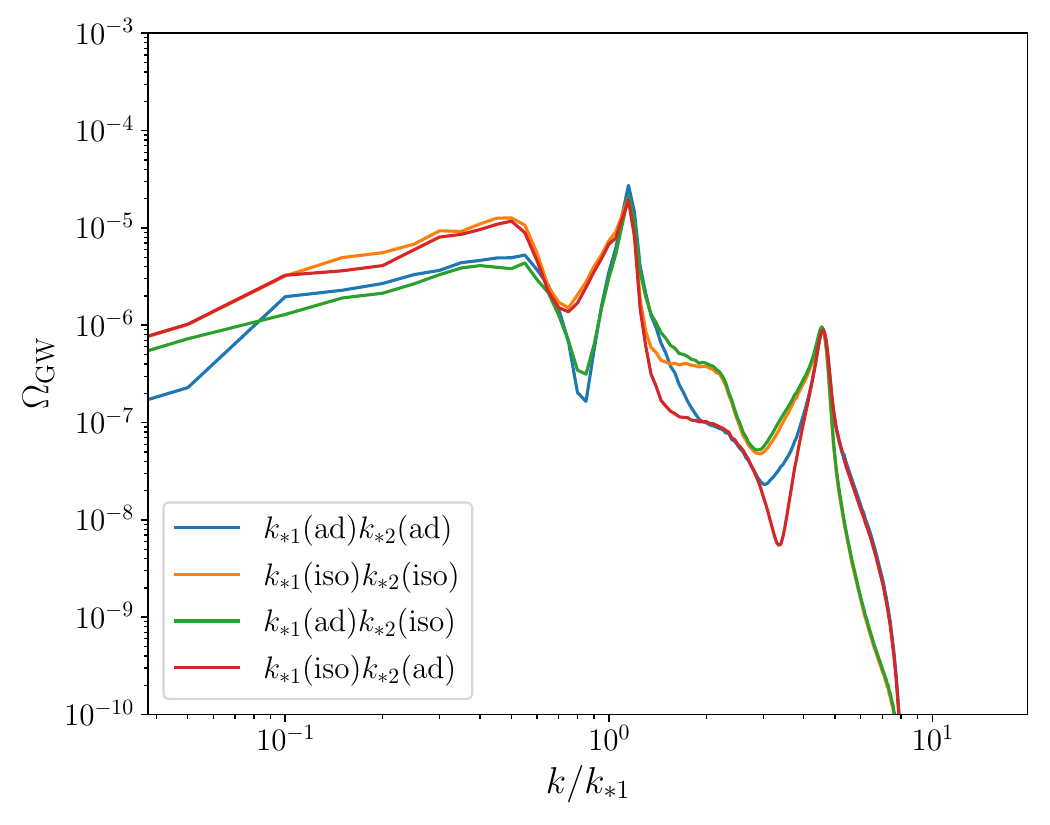}
    \caption{The GW energy spectra with different initial conditions. All power spectra of initial curvature perturbations are composed of two peaks of width $\Delta=1/30$ located at $k_{*1}$ and $k_{*2}$. In the left panel, $k_{*2}=3k_{*1}$, while in the right panel, $k_{*2} = 4k_{*1}$. ``ad'' and ``iso'' denote adiabatic and isocurvature perturbation, respectively. }
    \label{fig:compare1}
\end{figure}
To be specific, for the left panel of Fig.~\ref{fig:multi}, we adopt $B_1=B_2=100$ in Eq.~\eqref{eq:psdelta} to compute representative examples in isocurvature case; for the right panel, we choose $A=0.01$ in Eq.~\eqref{eq:psad} for adiabatic perturbation and $B=100$ in Eq.~\eqref{eq:psiso} for isocurvature perturbation for numerical evaluation in mixed initial condition case. To investigate whether different initial conditions can be distinguished via the GW energy spectra, we plot in Fig.~\ref{fig:compare1} the spectra induced by a curvature power spectrum featuring two peaks at $k_{*1}$ and $k_{*2}$. Imposing equal amplitudes for the first and last peaks, we find the amplitude of the additional intermediate peak to be largely unchanged. Thus, the initial conditions can be distinguished solely by the precise spectral shape around the peaks, a point also discussed in Ref.~\cite{Domenech:2021and}.

\section{Simulation results with energy transfer}\label{sec:pbh}
An early PBH-dominated era provides a viable mechanism for cosmic reheating~\cite{Lidsey:2001nj, Hidalgo:2011fj}, rendering the investigation of its observable imprints especially relevant. Additionally, certain non-topological solitons could also induce an early matter-dominated (eMD) era, potentially generating substantial gravitational wave signals~\cite{White:2021hwi}. In this section, we begin by examining SIGWs in the pure isocurvature case, and then extend the analysis to mixed initial conditions.

\subsection{SIGW from eMD isocurvature}
The SIGW signal induced by PBH isocurvature perturbations was initially studied and computed in Refs.~\cite{Papanikolaou:2020qtd, Domenech:2020ssp}. After their formation, PBHs are randomly distributed in space, following Poisson statistics to a very good approximation. As a result, the power spectrum of PBH density perturbations takes the form
\begin{align}
    \langle \delta\rho_{\mathrm{PBH}}(k)\delta\rho_{\mathrm{PBH}}(q) \rangle = \frac{4\pi}{3}\left(\frac{d}{a}\right)^3 \rho_{\mathrm{PBH}}^2 \delta(k+q),
\end{align}
where $k,q$ are the comoving wavenumber, and $d$ denotes the mean separation of PBHs, which can be expressed by 
\begin{align}
    d = \frac{a}{a_f}\left(\frac{3M_{\mathrm{PBH,f}}}{4\pi \rho_{\mathrm{PBH,f}}}\right)^{\frac{1}{3}},
\end{align}
the subindex $f$ means at the time of PBH formation, and $M_{\mathrm{PBH,f}}$ is the mass of PBH. The smoothed spectrum above is valid only on coarse-grained scales larger than the comoving ultraviolet cutoff scale, which is defined as
\begin{align}
    k_{\mathrm{UV}} \equiv \frac{a}{d} = a_fH_f\beta^{\frac13}\gamma^{-\frac13},
\end{align}
where $\gamma\sim0.2$ in the radiation-dominated (RD) era, $H_f$ is the Hubble parameter at the time of PBH formation, and $\beta$ denotes the fraction of PBHs at formation. In the lattice simulation, we define and input a dimensionless quantity $\tilde{k}_{\mathrm{UV}} = k_{\mathrm{UV}}/k_c$, where $k_c$ denotes some characteristic scale. Assuming that during the early radiation-dominated era (eRD), the energy density of PBHs is negligible compared to radiation at the time of PBH formation, and that initial density fluctuations on scales larger than $k_{\mathrm{UV}}^{-1}$ can be neglected, which means $\delta\rho_{\mathrm{PBH},f}+\delta\rho_{r,f}=0$.\footnote{Initial density fluctuations arise when adiabatic perturbations are considered. In contrast, isocurvature perturbations emerge following PBH formation, typically after the peak-like adiabatic perturbation has entered the horizon. Therefore, within linear perturbation theory, the adiabatic and isocurvature modes could be treated independently.} Therefore, the initial isocurvature perturbation is given by
\begin{align}
    S \simeq \frac{\delta\rho_{\mathrm{PBH,f}}}{\rho_{\mathrm{PBH,f}}},
\end{align}
which leads to the dimensionless initial isocurvature power spectrum
\begin{align}\label{eq:pspbhiso}
    \mathcal{P}_S(k) = \frac{2}{3\pi}\left(\frac{k}{k_{\mathrm{UV}}}\right)^3\Theta(k_{\mathrm{UV}}-k).
\end{align}
At early times following PBH formation, Hawking radiation remains negligible; hence, the initial conditions corresponding to the no-energy-transfer scenario may be adopted. The energy-transfer term can be defined as
\begin{align}
    aQ \equiv \rho_m a\Gamma,
\end{align}
where the decay rate is defined as
\begin{align}\label{eq:Gamma}
    \Gamma \equiv - \frac{\mathrm{d}\ln M_{\mathrm{PBH}}}{\mathrm{d}t},
\end{align}
with $t$ denoting the cosmic time. Consequently, $\delta Q = \delta\rho_m \Gamma$. The Hawking radiation decreases the PBH mass by~\cite{Kim:1999iv, Hooper:2019gtx}
\begin{align}
    \frac{\mathrm{d}M_{\mathrm{PBH}}}{\mathrm{d}t} = - \frac{A M_{\mathrm{pl}}^4}{M_{\mathrm{PBH}}^2},
\end{align}
where $A=3.8\pi g(T_{\mathrm{PBH}})/480$, and in our scenario $g(T_{\mathrm{PBH}})\simeq 108$. After integration, one can obtain the PBH mass
\begin{align}\label{eq:mpbh}
    M_{\mathrm{PBH}}(t) \simeq M_{\mathrm{PBH,f}}\left(1 - \frac{t}{t_{\mathrm{eva}}}\right)^{\frac13},
\end{align}
where $t_{\mathrm{eva}}$ describes the time of PBH evaporation, and we assume that $t_{\mathrm{eva}} \gg t_f$, thus,
\begin{align}
    t_{\mathrm{eva}} \simeq \frac{M_{\mathrm{PBH,f}}^3}{3AM_{\mathrm{pl}}^4},
\end{align}
then bringing Eq.~\eqref{eq:mpbh} into Eq.~\eqref{eq:Gamma}, one arrives at
\begin{align}
    \Gamma = \frac{1}{3(t_{\mathrm{eva}} - t )}.
\end{align}
Under the convention $a(\tau_i) = 1$, the conformal time roughly at matter-radiation equality, $\tau_*$, appeared in Eq.~\eqref{eq:ana_a}, is determined solely by the initial time $\tau_i$ and the initial abundance $\beta$\footnote{This $\tau_*$ will only be input as a initial parameter, the real $\tau_*$ can be obtained in the simulation.}
\begin{align}
    \tau_* = \frac{\tau_i}{\sqrt{\frac{1}{1-\beta}} - 1}.
\end{align}

\begin{figure}
    \centering
    \includegraphics[width=0.49\linewidth]{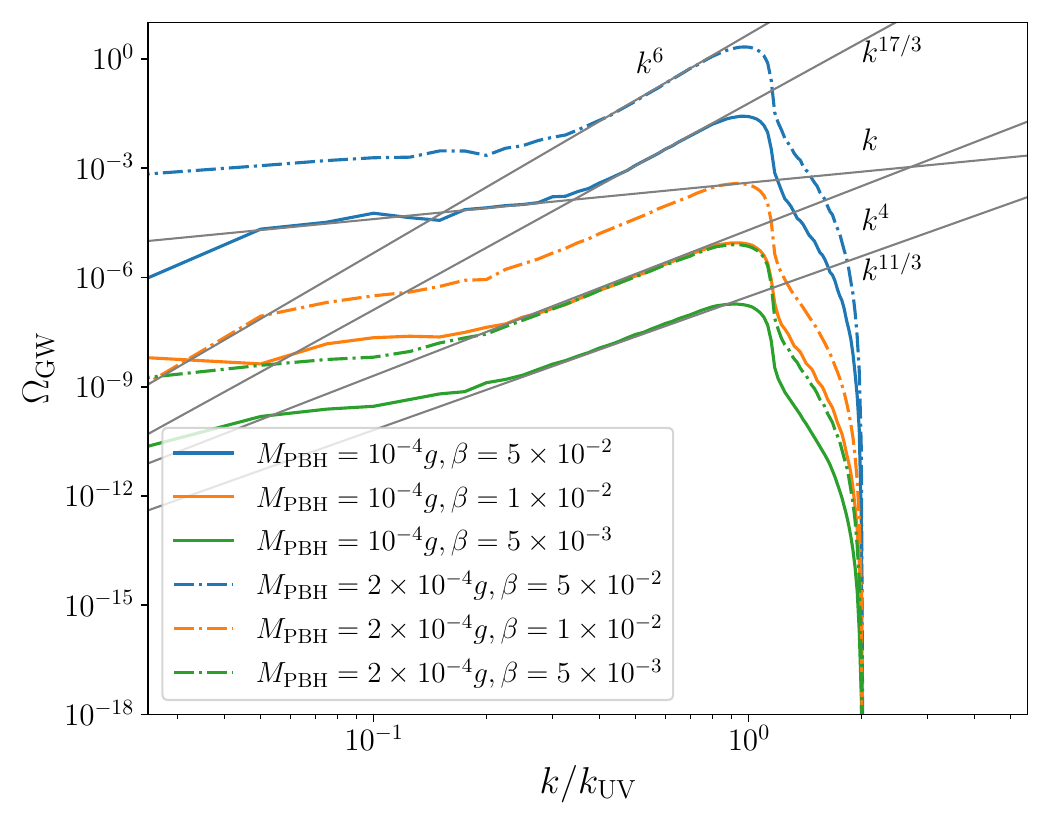}
    \includegraphics[width=0.49\linewidth]{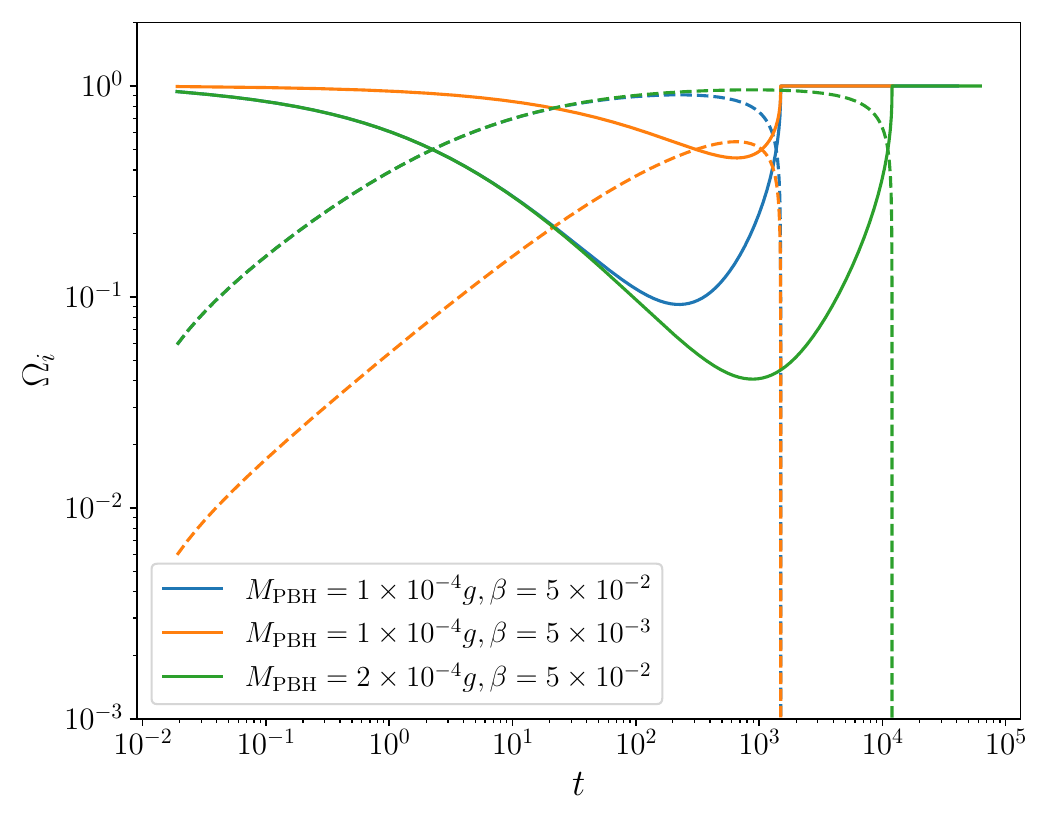}
    \caption{\textit{Left}: The GWs energy spectra sourced by PBH isocurvature perturbations. We have set $\tilde{k}_{\mathrm{UV}} = 40$ in our simulations. \textit{Right}: The energy density fraction $\Omega_i = \rho_{i=m,r}/(\rho_m+\rho_r)$ of PBH and radiation. The dashed lines denote $\Omega_m$, while the solid line describes $\Omega_r$. $t$ is the cosmic time. The gray lines are (fitting) reference lines indicating specific power-law scalings. The gray line with $k^{11/3}$ and $k$ scalings are the analytical solutions derived in Ref.~\cite{Domenech:2020ssp}.}
    \label{fig:pbhiso}
\end{figure}

\begin{figure}
    \centering
    \includegraphics[width=0.49\linewidth]{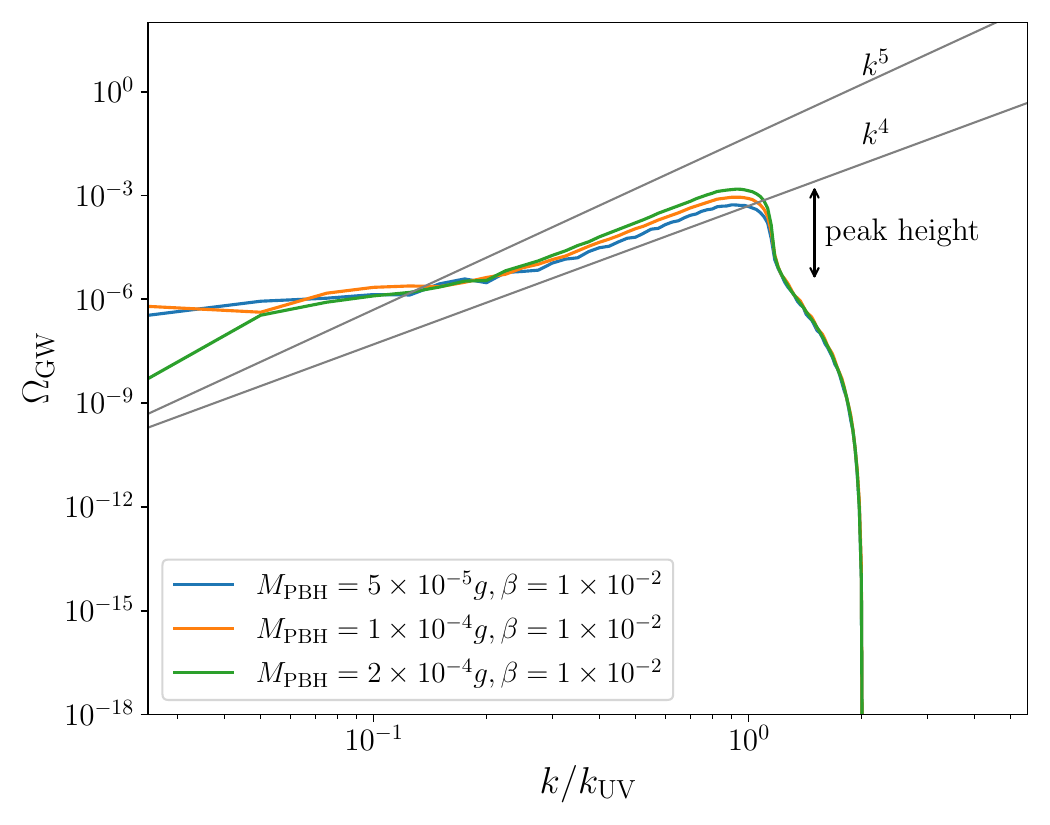}
    \includegraphics[width=0.49\linewidth]{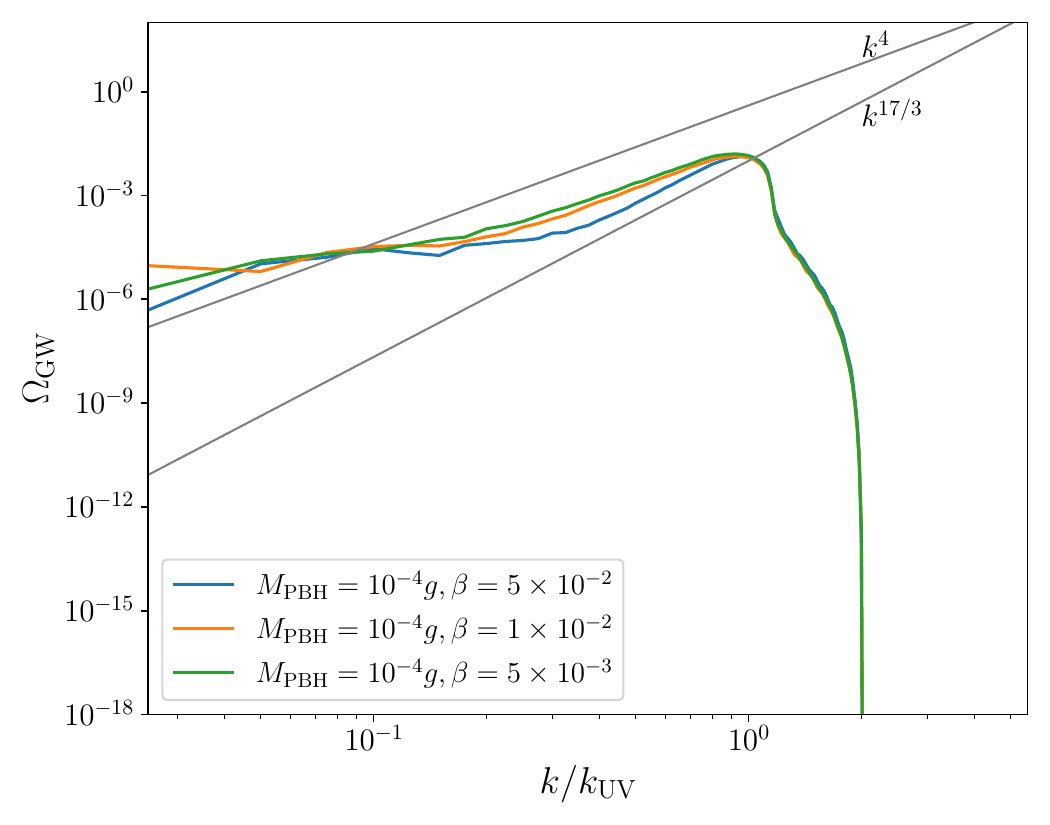}
    \caption{\textit{Left}: The GWs energy spectra sourced by PBH isocurvature perturbations for different PBH masses. \textit{Right}: The GWs energy spectra sourced by PBH isocurvature perturbations for different PBH abundances. In both figures, we have multiplied a constant to make their ultraviolet part the same, and the gray lines are (fitting) reference lines indicating specific power-law scalings.}
    \label{fig:compare}
\end{figure}

As illustrative examples, we consider PBH masses of $M_{\mathrm{PBH}} = 10^{-4}g$ and $2\times10^{-4}g$, and initial abundances $\beta = 5\times10^{-2}, 10^{-2}, 5\times10^{-3}$ in our simulations\footnote{For larger PBH masses or smaller initial abundances, the values of $t_{\mathrm{eva}}$ or $\tau_*$ in the simulation become very large. Simulating such cases is very challenging, which may require switching to a logarithmic time evolution scheme for the equations.}, as shown in Fig.~\ref{fig:pbhiso}. Our simulation has verified the $k$-scaling reported in Ref.~\cite{Domenech:2020ssp}. However, while the near-peak power law in that work is $k^{11/3}$, our results indicate a possible dependence on the PBH decay time and the duration of the PBH-dominated era. It can be seen that a larger initial abundance or a larger PBH mass leads to a steeper power-law behavior near the peak, as illustrated in Fig.~\ref{fig:compare}. An interesting finding is that the peak height (depicted in Fig.~\ref{fig:compare}) is dictated solely by the PBH mass, while the location of the break in the power-law scaling is governed only by the initial abundance of PBHs. This behavior may also originate from the different decay times of PBHs and the duration of the PBH-dominated era, an interpretation that warrants further analytical and numerical investigation in future work.

\begin{figure}
    \centering
    \includegraphics[width=0.49\linewidth]{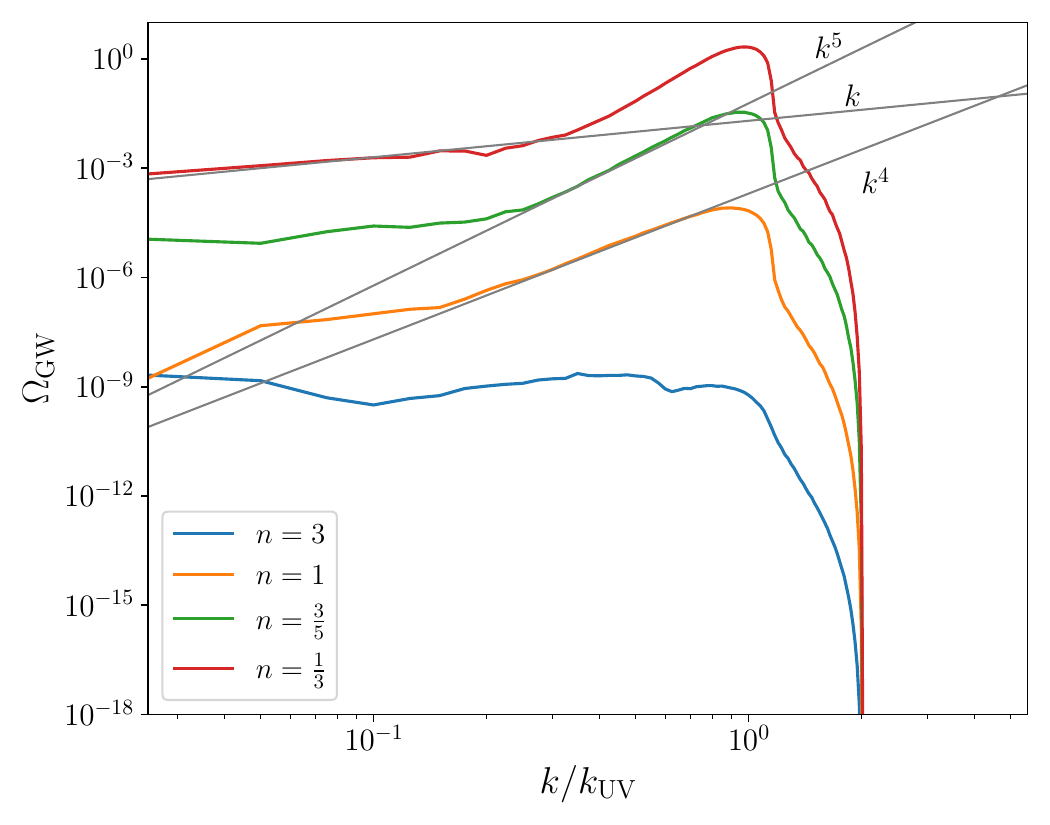}
    \includegraphics[width=0.49\linewidth]{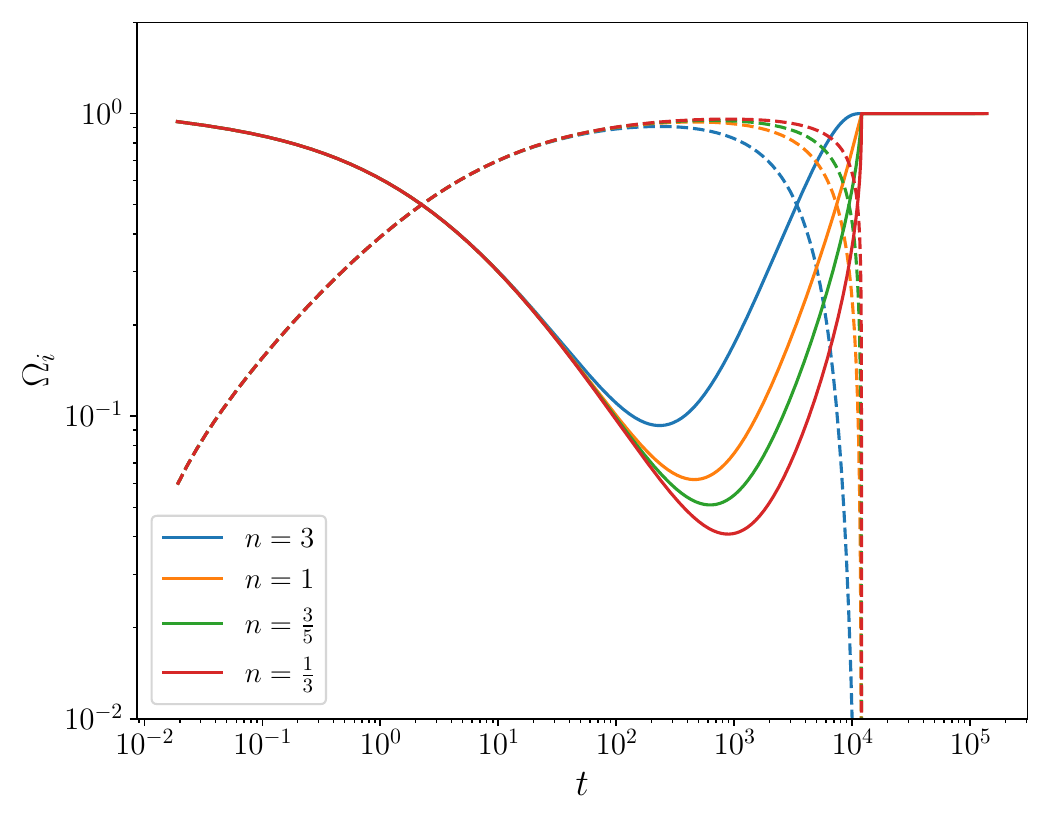}
    \caption{\textit{Left}: The GWs energy spectra sourced by Q-balls isocurvature perturbations. We have set $\tilde{k}_{\mathrm{UV}} = 40$ in our simulations. \textit{Right}: The energy density fraction $\Omega_i = \rho_{i=m,r}/(\rho_m+\rho_r)$ of Q-balls and radiation. The dashed lines denote $\Omega_m$, while the solid line describes $\Omega_r$. $t$ is the cosmic time. The initial abundance is set as $\beta = 5\times 10^{-2}$. The gray lines are (fitting) reference lines indicating specific power-law scalings.}
    \label{fig:qball}
\end{figure}

The dynamics of non-topological solitons, such as Q-balls, are closely analogous; the primary distinction lies in their decay rates
\begin{align}\label{eq:decayrate}
    \Gamma = \frac{n}{t_{\mathrm{eva}} - t },
\end{align}
where the parameter $n$ characterizes the decay rate for different types of solitons~\cite{Pearce:2025ywc, Multamaki:1999an}, with the values $n = 3, 1, 3/5$ corresponding to thin-wall Q-balls, thick-wall Q-balls, and delayed Q-balls, respectively. Figure~\ref{fig:qball} demonstrates that a slower decay rate results in both a shallower power-law index near the peak and a reduced GW amplitude. Consequently, it is imperative to derive an analytical relation between the decay rate and the GW amplitude in order to break degeneracies and distinguish among different Q-ball models and PBH populations.

\subsection{SIGW from general initial conditions}
\begin{figure}
    \centering
    \includegraphics[width=0.49\linewidth]{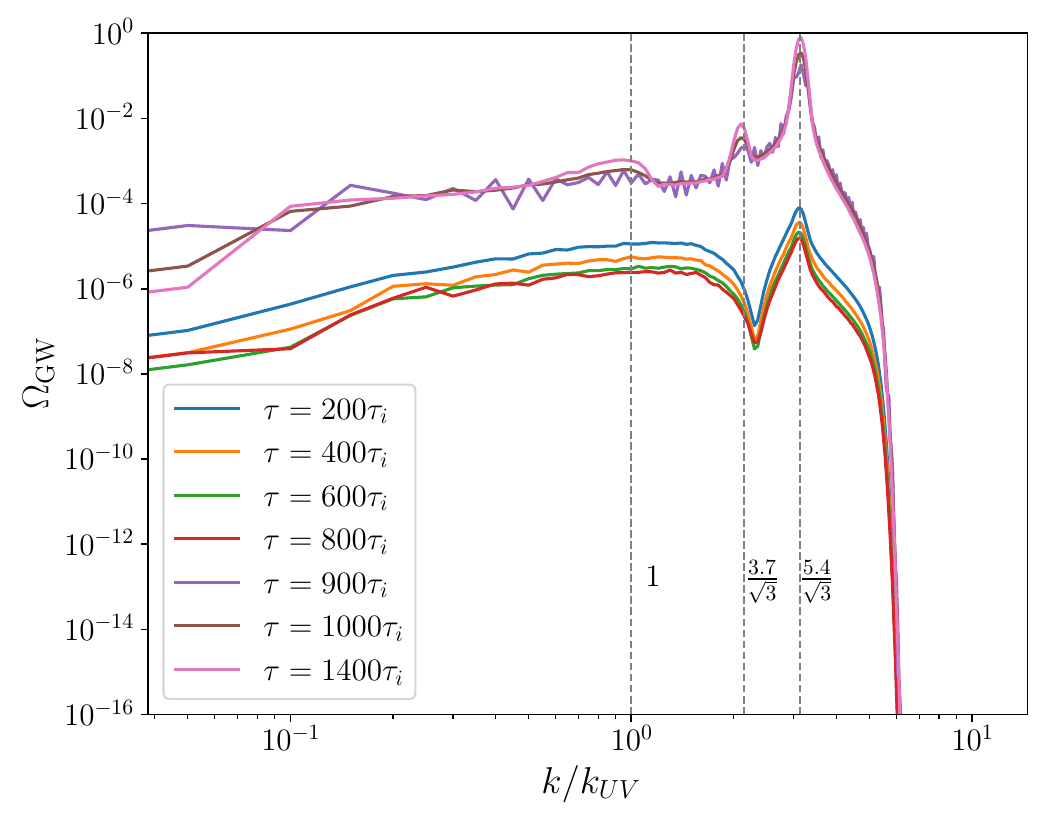}
    \includegraphics[width=0.49\linewidth]{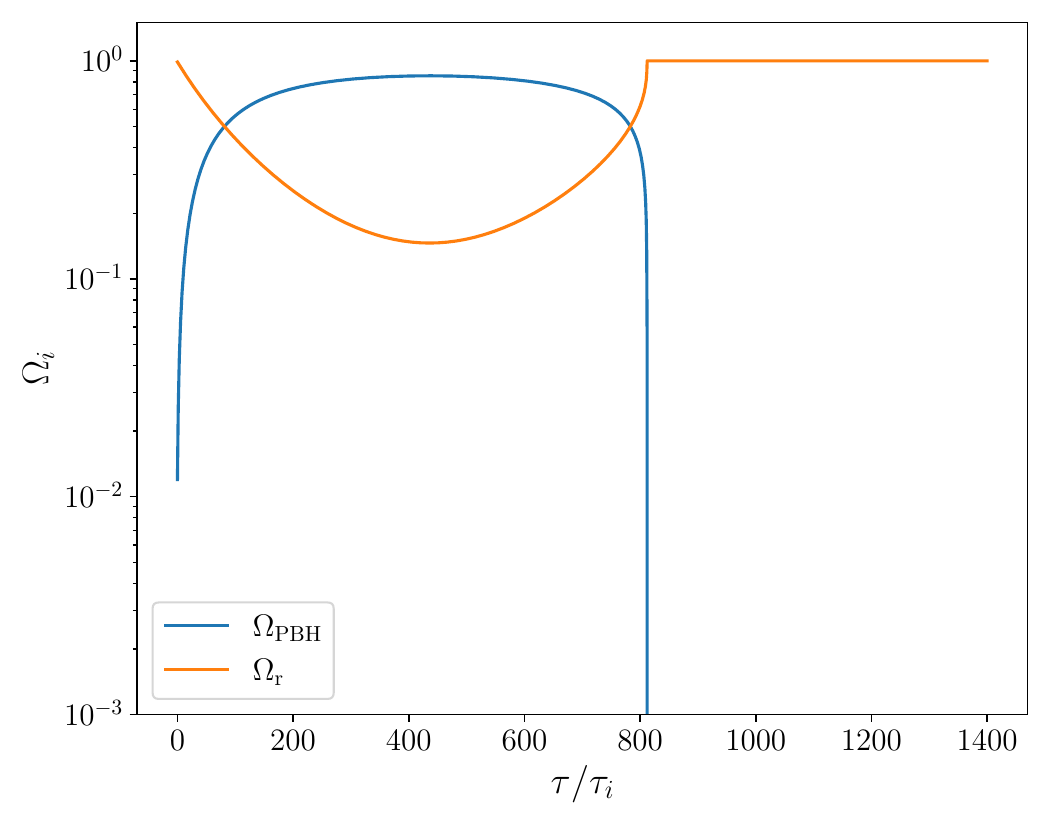}
    \caption{\textit{Left}: The GW energy spectrum sourced by the mixed initial condition at different time steps. The power spectrum of the initial adiabatic perturbation is the Gaussian bump power spectrum Eq.~\eqref{eq:psad} with $A=0.01, \Delta=1/30$ and $\tilde{k}_*\equiv k_*/k_c=2.7\tilde{k}_{\mathrm{UV}} = 54$, while the initial isocurvature power spectrum is Eq.~\eqref{eq:pspbhiso} with $\tilde{k}_{\mathrm{UV}} = 20$. $\tau_i$ is the initial comoving time. \textit{Right}: The energy density fraction of PBH and radiation. The mass of PBH is $M_{\mathrm{PBH}} = 2\times 10^{-4}g$ and the evaporation time is roughly $812\tau_i$.}
    \label{fig:pbhmix}
\end{figure}

A full treatment of general initial conditions is considerably more complex; here, we provide only a brief extension of the PBH-dominated case. Strictly speaking, isocurvature perturbations only emerge subsequent to PBH formation, typically well after the peak of the adiabatic perturbation has entered the horizon. Thus, one ought to initialize the adiabatic and isocurvature fluctuations at distinct times. However, the solutions~\eqref{eq:Sols} and \eqref{eq:solphi} are also attractors in the subhorizon regime, as noted in Ref.~\cite{Lozanov:2023aez}. Therefore, minor variations in the initial conditions should have a negligible impact if PBHs form sufficiently before radiation-PBH equality; thus, as an illustrative example, we simply superimpose the initial adiabatic and isocurvature contributions as the initial condition. Assuming that PBHs are formed from a peak-like adiabatic perturbation centered at $k_*$, the ultraviolet cutoff scale $k_{\mathrm{UV}}$ can be related to $k_*$ via the identification $k_* = \mathcal{H}_f$
\begin{align}
    k_{\mathrm{UV}} = \gamma^{-\frac13}\beta^{\frac13}k_*.
\end{align}
For $\beta = 10^{-2}$ and $\gamma=0.2$, one will obtain $k_*\simeq 2.7k_{\mathrm{UV}}$, which is shown in Fig.~\ref{fig:pbhmix}. As shown, the gravitational waves are initially generated by the adiabatic perturbation, as depicted by the blue line, and are subsequently sourced by the isocurvature perturbation. The rapid transition from the eMD to the RD era amplifies the GW signal, leading to a larger peak amplitude of the adiabatic part. This enhancement indicates that a general initial condition must be considered, rather than assuming a simple linear superposition of the adiabatic and isocurvature components. The middle peak corresponds to the peak structure discussed in Subsection~\ref{sec:peak}. By using \eqref{eq:peak}, one can find there will be three peaks located at $k_{UV},3.7/\sqrt{3}k_{UV}, 5.4/\sqrt{3}k_{UV}$.

\section{Discussion and conclusions}\label{sec:condis}
SIGWs represent a promising source of gravitational radiation, carrying valuable information about early-universe physics. With the advent of gravitational wave astronomy, accurate computation of the resulting GW energy spectra has become increasingly important. While numerous studies have examined SIGWs, most of them have focused primarily on adiabatic initial conditions. This study generalizes the approach of Ref.~\cite{Zeng:2025cer} to account for isocurvature initial conditions. With lattice simulations, it becomes feasible to compute SIGWs generated from mixed initial perturbations.

To validate our method, we first benchmark our simulation against semi-analytical results for pure isocurvature initial conditions~\cite{Domenech:2021and}, finding excellent agreement. However, when perturbations enter the horizon near the matter-radiation equality time, a discrepancy in amplitude emerges between the two approaches, suggesting the non-negligible nonlinear correction to the semi-analytical treatment. We then extend the treatment of multi-peak structures~\cite{Cai:2019amo, Zeng:2024ovg} to mixed initial conditions. Our results show that the number and locations of the peaks remain the same as in the purely adiabatic case. The primary distinction is the presence of a scale-dependent suppression factor affecting the isocurvature contribution.

The inclusion of the energy transfer term enables a more realistic treatment of the matter-radiation system in eMD scenarios.  We find that a larger initial PBH abundance or mass results in a steeper power-law spectrum near the peak. In contrast, for non-topological solitons (Q-balls), a faster decay rate (corresponding to a smaller $n$ in Eq.~\eqref{eq:decayrate}) yields a higher amplitude and a steeper slope near the peak. While the present study considers PBHs with very small masses, our results are directly relevant for guiding future analytical and numerical work. It is important to stress that our lattice simulation solves the perturbative equations of motion. Therefore, the present analysis does not account for phenomena like the nonlinear cutoff scale~\cite{Assadullahi:2009nf} or dynamics that lie outside the perturbative regime. Addressing this regime likely requires full numerical-relativity simulations or, at a minimum, the inclusion of the acoustic gravitational waves discussed in Refs.~\cite{Zeng:2025law, Ning:2025ogq, Ning:2025yvj}.

Several important aspects warrant further investigation. As noted in Ref.~\cite{Domenech:2021and}, isocurvature perturbations are intrinsically highly non-Gaussian due to the strongly non-Gaussian nature of the density contrast $\delta$. It is therefore important to systematically examine non-Gaussian effects in isocurvature perturbations, as well as their influence on adiabatic perturbations that also exhibit non-Gaussianity. Future studies of the eMD scenario should address the effects of PBH accretion~\cite{Allahverdi:2025btx} and the effect of clustering~\cite{He:2024luf, Papanikolaou:2024kjb}. In addition, direct simulations of PBHs with larger masses are also crucial. A detailed study of these aspects will be pursued in future work.

\acknowledgments
This work is supported by the National Key Research and Development Program of China Grant  No. 2021YFC2203004, No. 2021YFA0718304, and No. 2020YFC2201501, 
the National Natural Science Foundation of China Grants No. 12422502, No. 12105344, No. 12235019, No. 12047503, No. 12073088, No. 11821505, No. 11991052, No. 12447101, and No. 11947302,
and the China Manned Space Program Grant No. CMS-CSST-2025-A01.

\appendix

\section{Semi-analytical formulas in isocurvature case}\label{app:iso}
In this appendix, we will mainly follow the results of Ref.~\cite{Domenech:2021and}. Assuming the isocurvature perturbations are Gaussian, the energy spectrum of induced GWs at time $\tau_c$ is given by
\begin{align}
    \Omega_{\mathrm{GW}}(k,\tau_c) = \frac{2}{3}\int_{0}^{\infty} \mathrm{d}v \int_{|1-v|}^{1+v}\mathrm{d}u \left(\frac{4v^2 -(1-u^2+v^2)^2}{4uv}\right)^2\overline{I^2(k, \tau_c, u, v)}\mathcal{P}_S(ku)\mathcal{P}_S(kv),
\end{align}
where the oscillation-averaged kernel function is
\begin{align}
    \overline{I^2(k, \tau_c, u, v)} \simeq \frac{1}{2}(I^2_c(k, \tau_c\rightarrow\infty, u, v) + I^2_s(k, \tau_c\rightarrow\infty, u, v)),  
\end{align}
and the specific forms are
\begin{align}
    I_c(k, \tau_c\rightarrow\infty, u, v) &= \frac{9}{32u^4v^4\kappa^2}\Bigg{\{} -3u^2v^2 + (-3+u^2)(-3+u^2+2v^2)\ln \left|1-\frac{u^2}{3}\right| \nonumber\\
    &+(-3+v^2)(-3+v^2+2u^2)\ln\left|1-\frac{v^2}{3}\right|\nonumber\\
    &-\frac{1}{2}(-3+v^2+u^2)^2\ln\left[\left|1-\frac{(u+v)^2}{3}\right|\left|1-\frac{(u-v)^2}{3}\right|\right]\Bigg{\}}, \\
    I_s(k, \tau_c\rightarrow\infty, u, v) &= \frac{9\pi}{32u^4v^4\kappa^2}\Bigg{\{} 9 - 6v^2 -6u^2 + 2u^2v^2 \nonumber\\
    &+(3-u^2)(-3+u^2+2v^2)\Theta\left(1-\frac{u}{\sqrt{3}}\right) \nonumber\\
    &+(3-v^2)(-3+v^2+2u^2)\Theta\left(1-\frac{v}{\sqrt{3}}\right) \nonumber\\
    &+\frac{1}{2}(-3+v^2+u^2)^2\left[\Theta\left(1-\frac{u+v}{\sqrt{3}}\right) + \Theta\left(1+\frac{u-v}{\sqrt{3}}\right) \right] \Bigg{\}} ,
\end{align}
where $\kappa=k/k_{\mathrm{eq}}$, and $k_{\mathrm{eq}} = \mathcal{H}_{\mathrm{eq}}$.

\section{The effect of relative velocity}\label{app:vrel}
\begin{figure}[htbp]
    \centering
    \includegraphics[width=0.49\linewidth]{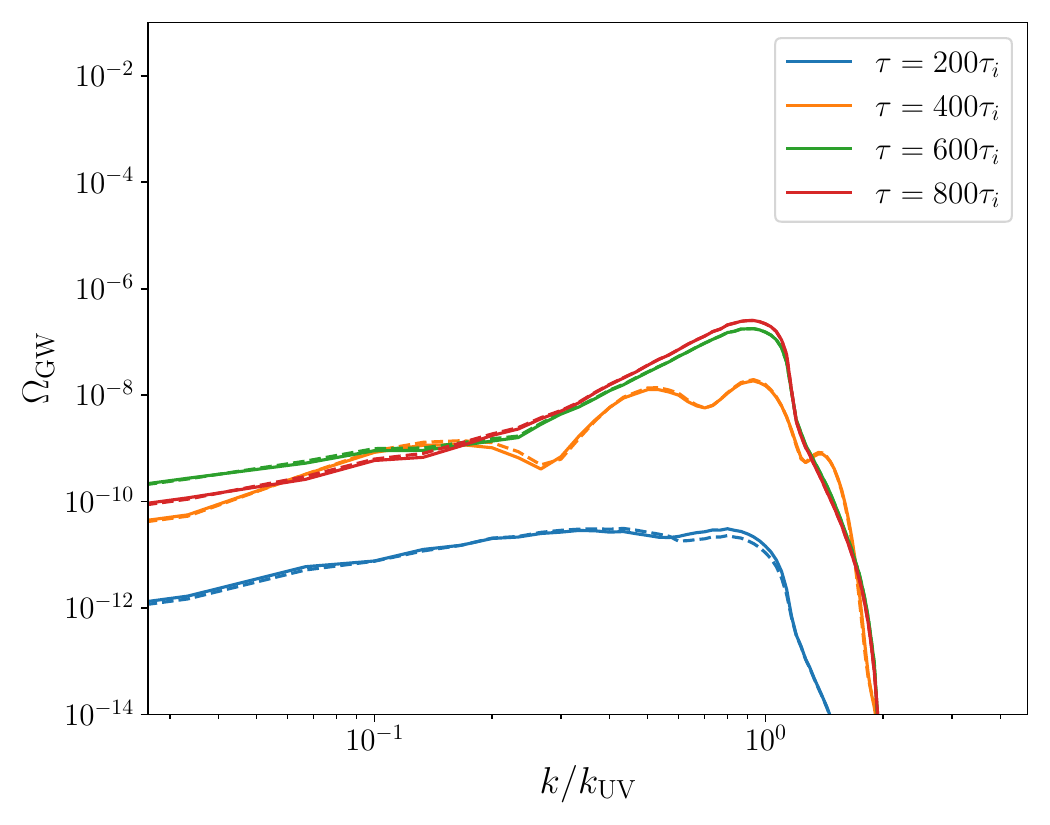}
    \includegraphics[width=0.49\linewidth]{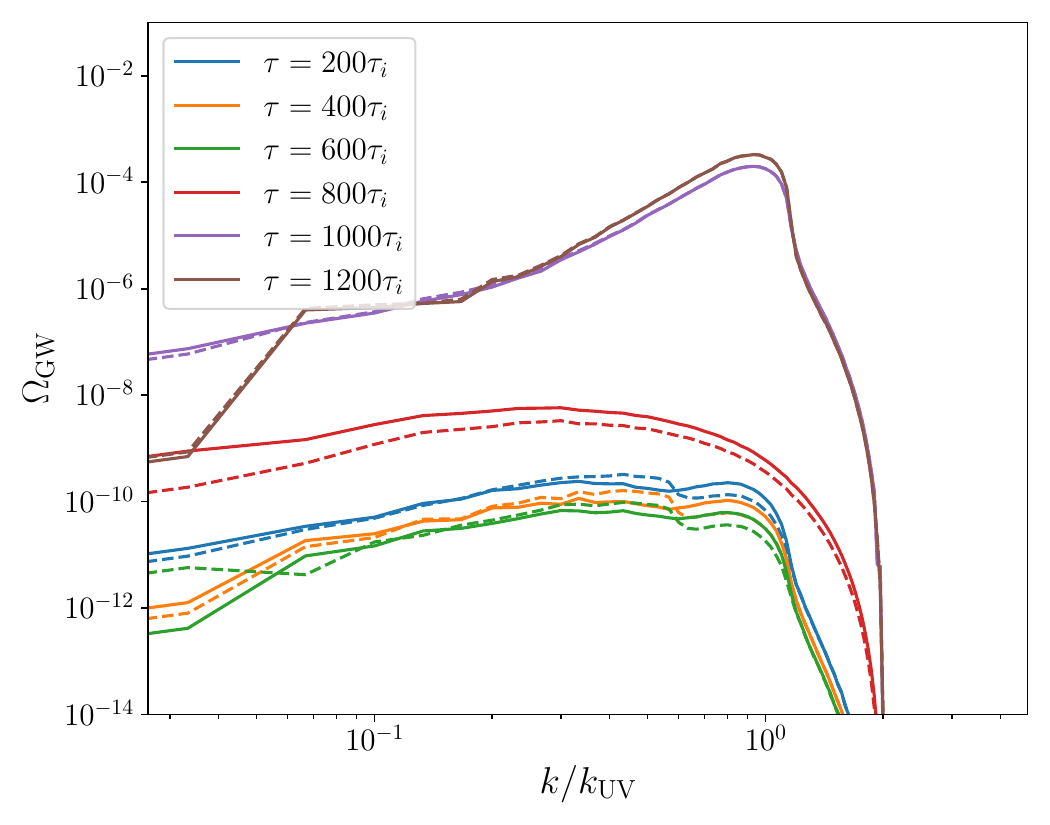}
    \caption{The GW energy spectra at different times. \textit{Left}: $M_{\mathrm{PBH}} = 10^{-4}g$, $\beta = 10^{-2}$. \textit{Right}: $M_{\mathrm{PBH}} = 2\times10^{-4}g$, $\beta = 10^{-2}$. We have also tested different $\beta$, and they show similar behaviors. The solid lines denote the results with relative velocity terms, while the dashed lines describe the results without relative velocity terms.}
    \label{fig:vrel}
\end{figure}

In this section, we tested the effect of the relative velocity. We observe that the effects are more prominent in the early stage of the scenario but are quickly overwhelmed by other contributions in the later stage, as shown in Fig.~\ref{fig:vrel}.

\bibliographystyle{JHEP}
\bibliography{ref}

\end{document}